\documentclass[%
 reprint,
 amsmath,amssymb,
 aps,
]{revtex4-2}

\usepackage{graphicx}
\usepackage{dcolumn}
\usepackage{bm}
\usepackage{xcolor}
\usepackage{algorithm} 
\usepackage{algpseudocode} 
\usepackage{tabularx}

\begin{document}


\title{Analysis of the autocorrelation function for time series with higher-order temporal correlations: An exponential case}

\author{Min-ho Yu}
\affiliation{Department of Physics, The Catholic University of Korea, Bucheon, Republic of Korea}

\author{Hang-Hyun Jo}
\email{h2jo@catholic.ac.kr}
\affiliation{Department of Physics, The Catholic University of Korea, Bucheon, Republic of Korea}

\date{\today}

\begin{abstract}
Temporal correlations in the time series observed in various systems have been characterized by the autocorrelation function. Such correlations can be explained by heavy-tailed interevent time distributions as well as by correlations between interevent times. The latter is called higher-order temporal correlations, and they have been captured by the notion of bursts; a burst indicates a set of consecutive events that rapidly occur within a short time period and are separated from other bursts by long time intervals. The number of events in the burst is called a burst size. Some empirical analyses have shown that consecutive burst sizes are correlated with each other. To study the impact of such correlations on the autocorrelation function, we devise a model generating a time series with higher-order temporal correlations by employing the copula method. We successfully derive the analytical solution of the autocorrelation function of the model time series for arbitrary distributions of interevent times and burst sizes when consecutive burst sizes are correlated. For the demonstration of our analysis, we adopt exponential distributions of interevent times and burst sizes to understand how the correlations between consecutive burst sizes affect the decaying behavior of the autocorrelation function.
\end{abstract}

\maketitle

\section{Introduction}\label{sec:intro}

Time series observed in various natural and social systems have shown complex temporal patterns with long-term temporal correlations~\cite{Hurst1956Problem, Brunetti1984Analysis, Weissman1988Noise, Bak1987Selforganized, Mantegna1999Introduction, Kantelhardt2001Detecting, Kantz2004Nonlinear, Allegrini2009Spontaneous, Panzarasa2015Emergence, Karsai2018Bursty}. One of the most commonly used methods characterizing temporal correlations is the autocorrelation function (ACF)~\cite{Fano1950Shorttime, Kantelhardt2001Detecting}. The ACF for the time series $x(t)$ is defined as
\begin{align}
    A(t_{\rm d})\equiv \frac{ \langle x(t)x(t+t_{\rm d})\rangle- \langle x(t)\rangle^2}{ \langle x(t)^2\rangle- \langle x(t)\rangle^2},
    \label{eq:acf_define}
\end{align}
where $\langle\cdot\rangle$ is the time average taken over the entire period in the time series, and $t_{\rm d}$ is the time lag. For the time series with long-term correlations, one often finds an algebraically decaying behavior such that $A(t_{\rm d})\propto t_{\rm d}^{-\gamma}$, where $\gamma$ is the decay exponent.

The time series that is given as a sequence of event timings is called an event sequence. The empirical event sequences have been characterized in terms of the interevent time (IET) which is the time interval between two consecutive events and denoted by $\tau$~\cite{Barabasi2005Origin, Karsai2018Bursty, Jo2023Bursty}. In particular, heavy-tailed IET distributions, $P(\tau)$, imply long-term memory effects in the time series as the memoryless, Poisson processes lead to exponential IET distributions. Recently, the notion of bursty trains was introduced to capture correlations between IETs that are missing in the IET distribution~\cite{Karsai2012Universal}. For a given timescale $\Delta t$, a bursty train or a burst is defined as a set of consecutive events where each pair of consecutive events in the burst are separated by an IET less than or equal to $\Delta t$, while events in different bursts are separated by IETs larger than $\Delta t$. The number of events in the burst is called a burst size, which is denoted by $b$. Empirical analyses have shown that the burst size distributions, $Q_{\Delta t}(b)$, are heavy-tailed for a wide range of $\Delta t$, implying the presence of correlations between arbitrarily many consecutive events~\cite{Karsai2012Universal, Karsai2012Correlated, Yasseri2012Dynamics, Jiang2013Calling, Kikas2013Bursty, Wang2015Temporal, Karsai2018Bursty, Jo2020Bursttree}. In addition, consecutive burst sizes turn out to be positively correlated with each other for a wide range of $\Delta t$ in several datasets from diverse backgrounds~\cite{Jo2020Bursttree}. It means that bigger (smaller) bursts tend to follow bigger (smaller) ones, which can be related to the bursty-get-burstier mechanism~\cite{Jo2017Modeling} that has been proposed to explain heavy-tailed burst size distributions. We call such empirical findings higher-order temporal correlations as they are correlations beyond the IET distributions~\cite{Birhanu2025Bursttree}.

Temporal correlations in event sequences, characterized by the ACF, can be explained by the heavy-tailed IET distribution as well as by correlations between IETs. The effects of the latter, higher-order temporal correlations on the ACF have been largely unexplored. For the simplest, Poisson process, it is obvious that $A(t_{\rm d})=\delta(t_{\rm d})$, where $\delta(\cdot)$ is a Dirac delta function. For the renewal process in which IETs are fully uncorrelated with each other, the power-law IET distribution with the exponent $\alpha$ has been analytically shown to lead to the decay exponent of the ACF as $\gamma=|\alpha-2|$ for $1<\alpha<3$~\cite{Lowen1993Fractal}. The cases with correlated IETs were also studied; when the IETs are power-law distributed with exponent $\alpha$ and each IET depends only on the previous IET in the time series, the scaling relation between $\alpha$ and $\gamma$ remains the same, but the ACF is elevated for the range of the small $t_{\rm d}$ due to the positive correlations between two consecutive IETs~\cite{Jo2019Analytically}. More recently, for the case where the correlations between IETs were implemented by a power-law burst size distribution with the exponent $\beta$, the relation between $\alpha$, $\beta$, and $\gamma$ has been analytically obtained by devising a discrete-time model considering only one timescale $\Delta t=1$ for defining bursts~\cite{Jo2024Temporal}. However, the effects of correlations between two consecutive burst sizes on the ACF have not been studied, which is the aim of our current work.

In this work, we extend the time series model proposed in Ref.~\cite{Jo2024Temporal} to incorporate the correlations between two consecutive burst sizes. In particular, the correlations between two consecutive burst sizes are implemented using the Farlie-Gumbel-Morgenstern (FGM) copula method~\cite{Schucany1978Correlation, Nelsen2006Introduction, Takeuchi2010Constructing, Cossette2013Multivariate, Takeuchi2020Constructing, Jo2022Copulabased}. The copula method enables us to write a joint probability distribution function of correlated variables in a tractable form~\cite{Nelsen2006Introduction}. Note that the FGM copula was employed in the previous works to implement correlations between two consecutive IETs~\cite{Jo2019Analytically, Jo2019Copulabased, Jo2023Copulabased}. Then we derive an analytical solution of the ACF as a function of correlations between two consecutive burst sizes for arbitrary IET and burst size distributions. To demonstrate the validity of our analysis, we consider exponential distributions of both IETs and burst sizes but with their lower bounds. We find that the positive correlation between two consecutive burst sizes elevates the ACF curve. Finally, we perform numerical simulations to confirm our analytical results.

\section{Model}\label{sec:model}

Following Ref.~\cite{Jo2024Temporal}, we consider a discrete-time model that generates a time series, $\left\{x(1),x(2),\ldots,x(T)\right\}$, where $T$ is the number of discrete times. We set $x(t)=1$ if an event occurs at a time $t$, $x(t)=0$ otherwise. By definition, the minimum interevent time (IET) is $\tau_{\rm min}=1$. As for the timescale defining bursts, we consider only $\Delta t=1$, implying that any maximal set of consecutively occurring events is detected as a burst. An isolated event can also make its own burst. The number of events in a burst is called a burst size, which is denoted by $b$. Considering the fact that each burst of size $b$ generates exactly $b-1$ IETs of $\tau=1$, we separate IETs of $\tau=1$ from those of $\tau\geq 2$. That is, we assume that the fraction of IETs of $\tau=1$ is solely determined by the burst size distribution $Q(b)$, while IETs of $\tau\geq 2$ are fully governed by a separate IET distribution $\psi(\tau)$. Note that $\psi(\tau)$ is normalized as $\sum_{\tau= 2}^{\infty}\psi(\tau)=1$. Then, the IET distribution for the entire range of $\tau\geq 1$ can be written as
\begin{align}
    P(\tau)=\left(1-\frac{1}{\langle b\rangle}\right)\delta(\tau,1)+\frac{1}{\langle b\rangle}[1-\delta(\tau,1)]\psi(\tau),
\end{align}
where $\delta(\cdot,\cdot)$ is a Kronecker delta and $\langle b\rangle$ is the average burst size for a given $Q(b)$~\cite{Jo2024Temporal}:
\begin{align}
    \langle b\rangle\equiv \sum_{b=1}^\infty bQ(b).
\end{align}
Therefore, the IET distribution $\psi(\tau)$ for $\tau\geq 2$ and the burst size distribution $Q(b)$ for $b\geq 1$ can be independent.

To incorporate correlations between two consecutive burst sizes into the model, we assume that the joint probability distribution function of two consecutive burst sizes, namely, $Q(b_i, b_{i+1})$, carries the information on such correlations. Then for a given burst size $b_i$, the next burst size $b_{i+1}$ can be drawn from the conditional distribution function $Q(b_{i+1}|b_i)=Q(b_i, b_{i+1})/Q(b_i)$. The precise form of $Q(b_i, b_{i+1})$ will be discussed later. Finally, we also assume that correlations between IETs exist only within each burst, hence IETs drawn from $\psi(\tau)$ are independent of each other and of burst sizes as well.

\begin{figure}[!t]
\centering
\includegraphics[width=\columnwidth]{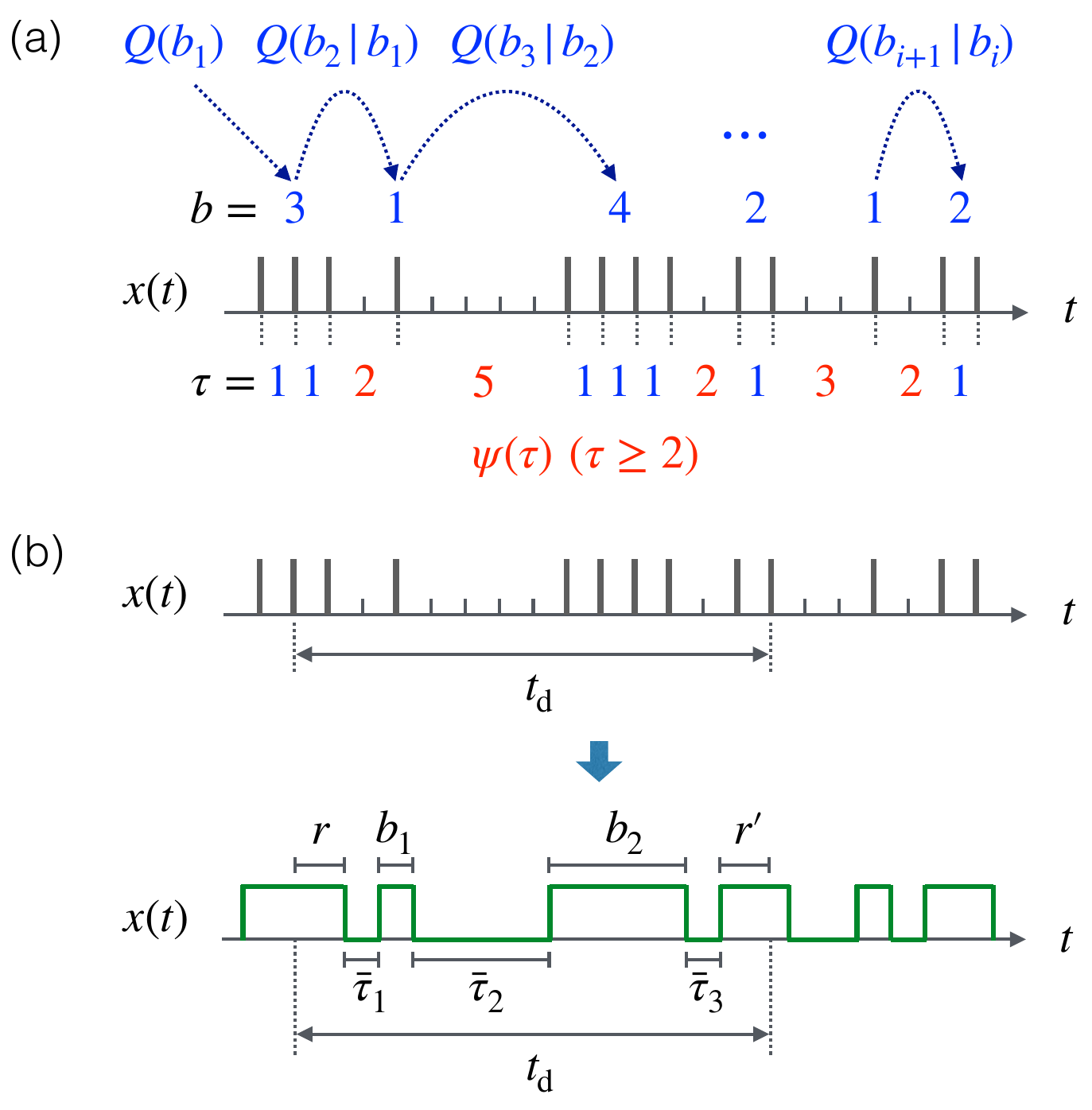}
\caption{(a) Schematic diagram of the model for generating an event sequence with given interevent time distribution $\psi(\tau)$ ($\tau\geq 2$) and burst size distribution $Q(b)$ ($b\geq 1$) when two successive burst sizes are correlated with each other via the conditional probability distribution $Q(b_{i+1}|b_i)$. (b) An example of the time lag $t_{\rm d}$ for the autocorrelation function in Eq.~\eqref{eq:acf_define}; the time interval of $t_{\rm d}$ can be seen as alternations of periods of $x(t)=1$ and $x(t)=0$. See the main text for definitions of symbols.
}
\label{fig:scheme}
\end{figure}

Let us define the model; see Fig.~\ref{fig:scheme}(a) for a schematic diagram of the model. We first draw a burst size $b_1$ from $Q(b)$ randomly and set $x(t)=1$ for $t=1,\ldots,b_1$. Then an IET $\tau_1$ is drawn from $\psi(\tau)$ randomly to set $x(t)=0$ for $t=b_1+1,\ldots,b_1+\tau_1-1$. The next burst size $b_2$ is drawn from $Q(b|b_1)$ and we set $x(t)=1$ for $t=b_1+\tau_1,\ldots,b_1+\tau_1+b_2-1$. Another IET $\tau_2$ is drawn from $\psi(\tau)$ and we set $x(t)=0$ for $t=b_1+\tau_1+b_2,\ldots,b_1+\tau_1+b_2+\tau_2-2$. This procedure is repeated until $t$ becomes $T$.

Although our model is defined in discrete times, we will assume continuous IETs and burst sizes to derive the autocorrelation function for analytical tractability. Accordingly, we also assume the joint probability distribution function $Q(b_i,b_{i+1})$ for continuous burst sizes. Precisely, we adopt the Farlie-Gumbel-Morgenstern (FGM) copula~\cite{Schucany1978Correlation, Nelsen2006Introduction} to write
\begin{align}
	Q(b_i, b_{i+1})=Q(b_i)Q(b_{i+1}) \left[1+\rho f(b_i)f(b_{i+1})\right],
	\label{eq:Q_bb}
\end{align}
where 
\begin{align}
    f(b) \equiv 2F(b)-1,\ F(b) \equiv \int_0^{b} {\rm d}b' Q(b').
\end{align}
Here $F(b)$ is the cumulative distribution function of burst sizes. The correlation parameter $\rho\in[-1,1]$ controls the correlation between burst sizes $b_i$ and $b_{i+1}$. Using Eq.~\eqref{eq:Q_bb} one can define the memory coefficient for bursts $M_b$~\cite{Jo2020Bursttree} as follows:
\begin{align}
    M_b\equiv \frac{\langle b_ib_{i+1}\rangle-\langle b\rangle^2}{\sigma_b^2},
    \label{eq:Mb_define}
\end{align}
where
\begin{align}
    \langle b_ib_{i+1}\rangle\equiv \int_0^{\infty} {\rm d}b_i\int_0^{\infty} {\rm d}b_{i+1}b_ib_{i+1}Q(b_i, b_{i+1}),
\end{align}
and $\sigma_b^2$ denotes the variance of burst sizes. $M_b$ has a value in the range of $[-1,1]$; the positive $M_b$ means that a big (small) burst tends to follow a big (small) one. The negative $M_b$ implies the opposite tendency, while $M_b\approx 0$ indicates no correlations between two consecutive burst sizes. Note that the memory coefficient was originally proposed to measure the correlation between two consecutive IETs~\cite{Goh2008Burstiness}. It turns out that $M_b$ in Eq.~\eqref{eq:Mb_define} is a linear function of $\rho$:
\begin{align}
    M_b=a\rho,\ a\equiv \frac{1}{\sigma_b^2}\left[\int_0^\infty {\rm d}b b  Q(b)f(b)\right]^2.
    \label{eq:Mb_arho}
\end{align}
Since $|\rho|\leq 1$, we get $|M_b|\leq a$. The upper bound of $a$ for any $Q(b)$ has been proven to be $1/3$~\cite{Schucany1978Correlation}.

In sum, our model has three inputs, namely, the IET distribution for $\tau\geq 2$, $\psi(\tau)$, the burst size distribution $Q(b)$, and the correlation parameter $\rho$ in the FGM copula for implementing $Q(b_{i+1}|b_i)$.

\section{Results}\label{sec:result}

\subsection{Analytical framework}\label{subsec:analysis}

We provide the analytical framework for deriving the autocorrelation function (ACF) in Eq.~\eqref{eq:acf_define} for the event sequence generated by our model introduced in Sec.~\ref{sec:model}. Since by definition $A(0)=1$, we consider $t_{\rm d}>0$ unless otherwise stated. If $n$ events occur in the model time series $\{x(1),\ldots,x(T)\}$, we get the event rate as
\begin{align}
    \lambda\equiv \langle x(t)\rangle = \frac{n}{T}.
\end{align}
We also get $\langle x(t)^2\rangle=\langle x(t)\rangle=\lambda$ because $x(t) \in \{ 0, 1 \}$. The term $\langle x(t)x(t+t_{\rm d})\rangle$ in Eq.~\eqref{eq:acf_define} is written as
\begin{align}
    \langle x(t)x(t+t_{\rm d})\rangle&=\Pr[x(t)=1\wedge x(t+t_{\rm d})=1]\nonumber\\
    &=\Pr[x(t)=1]\cdot \Pr[x(t+t_{\rm d})=1|x(t)=1]\nonumber\\
    &=\lambda Z(t_{\rm d}).
\end{align}
Here $Z(t_{\rm d})$ denotes the probability that $x(t+t_{\rm d})=1$ conditioned on $x(t)=1$. Thus, the ACF reads
\begin{align}
    A(t_{\rm d})=\frac{Z(t_{\rm d})-\lambda}{1-\lambda}.
    \label{eq:acf_simple}
\end{align} 

Any given time lag $t_{\rm d}$ can be seen as alternations of periods of $x(t)=1$ and $x(t)=0$, as depicted in Fig.~\ref{fig:scheme}(b). Each period of $x(t)=1$ is due to a burst, while that of $x(t)=0$ is due to an interevent time (IET) of $\tau\geq 2$. We denote by $k$ the number of IETs of $\tau\geq 2$ within the interval of $[t,t+t_{\rm d}]$. Then $Z(t_{\rm d})$ can be written as
\begin{align}
    Z(t_{\rm d})=\sum_{k=0}^\infty p_k(t_{\rm d}),
    \label{eq:Z_td}
\end{align}
where $p_k(t_{\rm d})$ is the probability that an event occurs at the time $t + t_{\rm d}$ after exactly $k$ IETs of $\tau\geq 2$, conditioned that an event occurs at the time $t$. The case with $k=0$ indicates that both events at times $t$ and $t+t_{\rm d}$ occur in the same burst, which happens for bursts of size $b\geq t_{\rm d}+1$. Thus one gets
\begin{align}
    p_0(t_{\rm d})=\frac{1}{\langle b\rangle} \sum_{b=t_{\rm d}+1}^\infty Q(b)(b-t_{\rm d}).
    \label{eq:p_0}
\end{align}

When $k\geq 1$, events at times $t$ and $t+t_{\rm d}$ belong to different bursts. The number of events since the time $t$ to the last event of the burst having the event at the time $t$ is denoted by $r$. When we count $r$, we exclude the event at time $t$, leading to $r\geq 0$. The probability distribution function of $r$ is given by
\begin{align}
    R(r)=\frac{1}{\langle b\rangle}\sum_{b=r+1}^\infty Q(b).
\end{align}
Next, $r'$ denotes the number of events that exist in the burst having the event at the time $t+t_{\rm d}$ and exist before or at the time $t+t_{\rm d}$. By definition, $r'\geq 1$. The probability of an event to occur at time $t+t_{\rm d}$ when the burst containing this event starts at time $t+t_{\rm d}+r'-1$ is denoted by $q(r')$, which is given by
\begin{align}
    q(r')=\sum_{b=r'}^\infty Q(b).
\end{align}
Then the time lag $t_{\rm d}$ is written by [Fig.~\ref{fig:scheme}(b)]
\begin{align}
    t_{\rm d} &= r+ \sum_{i=1}^k \tau_i+\sum_{i=1}^{k-1}(b_i-1)+r'-1\notag\\
    &= r+ \sum_{i=1}^k\bar\tau_i+\sum_{i=1}^{k-1}b_i+r',
\label{eq:td_define}
\end{align}
where a reduced IET has been defined as
\begin{align}
    \bar\tau_i \equiv \tau_i - 1
    \label{eq:bartau_define}
\end{align}
for convenience. Note that $\bar{\tau}_i \ge 1$ because $\tau_i\geq 2$. It is also remarkable that the event rate $\lambda$ is related to the average reduced IET and the average burst size as
\begin{align}
    \lambda= \langle x(t)\rangle = \frac{\langle b\rangle}{\langle\bar\tau\rangle+\langle b\rangle},
    \label{eq:lambda}
\end{align}
where
\begin{align}
    \langle\bar\tau\rangle\equiv \sum_{\bar\tau=1}^\infty \bar\tau\psi(\bar\tau).
\end{align}
Finally, $p_k(t_{\rm d})$ for $k\geq 1$ reads
\begin{align}
    p_k(t_{\rm d})=
    &\sum_{r=0}^\infty\left(\prod_{i=1}^k \sum_{\bar\tau_i=1}^\infty\right) \left(\prod_{i=1}^{k-1} \sum_{b_i=1}^\infty\right)\sum_{r'=1}^\infty \nonumber\\
    & \left[\prod_{i=1}^k \psi(\bar\tau_i)\right] P(r,b_1,\ldots,b_{k-1},r') 
    \nonumber\\
    &\times \delta\left(t_{\rm d},  
    r+\sum_{i=1}^k\bar\tau_i + \sum_{i=1}^{k-1}b_i+r'\right),
    \label{eq:p_k}
\end{align}
where the joint probability distribution function $P(r,b_1,\ldots,b_{k-1},r')$, or simply denoted by $P[k]$, will be explicitly written.

From now on, all variables for reduced IETs and burst sizes as well as $r$ and $r'$ are assumed to be real numbers. Thus, all distributions and probabilities are considered for their continuous variables. One gets
\begin{align}
    p_0(t_{\rm d})&=\frac{1}{\langle b\rangle} \int_{t_{\rm d}}^\infty {\rm d}b Q(b)(b-t_{\rm d}),
    \label{eq:p_0_cont}\\   
    R(r)&=\frac{1}{\langle b\rangle}\int_{r}^\infty {\rm d}b Q(b),
    \label{eq:R_cont_define}\\
    q(r')&= \int_{r'}^\infty {\rm d}b Q(b),
    \label{eq:q_cont_define}
\end{align}
and for $k\geq 1$
\begin{align}
    p_k(t_{\rm d})&=
    \int_0^\infty {\rm d}r \left(\prod_{i=1}^k \int_0^\infty {\rm d}\bar\tau_i\right) \left(\prod_{i=1}^{k-1} \int_0^\infty {\rm d}b_i\right) \int_0^\infty {\rm d}r' \nonumber\\
    &\left[\prod_{i=1}^k \psi(\bar\tau_i)\right] P[k]\, \delta\left(t_{\rm d}- 
    r- \sum_{i=1}^k\bar\tau_i-\sum_{i=1}^{k-1}b_i-r'\right),
    \label{eq:p_k_cont}
\end{align}
where $\delta(\cdot)$ is a Dirac delta. The product of reduced IET distributions $\psi(\bar\tau_i)$ in Eq.~\eqref{eq:p_k_cont} is due to the assumption that those IETs are independent of each other as well as of burst sizes. As for $P[k]$, we consider the correlation between two consecutive burst sizes, enabling us to write for $k=1$
\begin{align}
    P[1]=R(r)q(r'|r),
    \label{eq:P[1]_define}
\end{align}
and for $k\geq 2$
\begin{align}
    P[k]=R(r) Q(b_1|r)\left[\prod_{i=1}^{k-2}Q(b_{i+1}|b_i)\right]q(r'|b_{k-1}).
    \label{eq:P[k]_define}
\end{align}
The conditional distribution functions $Q(\cdot|\cdot)$ and conditional probabilities $q(\cdot|\cdot)$ will be derived soon. For this, we first consider the case with $k\geq 2$ and then return to the case with $k=1$. 

For $k\geq 2$, let us derive the conditional distribution function $Q(b_{i+1}|b_i)$ using the Farlie-Gumbel-Morgenstern (FGM) copula in Eq.~\eqref{eq:Q_bb}, which is given as
\begin{align}
	Q(b_{i+1}|b_i)=Q(b_{i+1}) \left[1+\rho f(b_i)f(b_{i+1})\right].
	\label{eq:Q_b|b}
\end{align}
As for $Q(b_1|r)$ and $q(r'|b_{k-1})$, note that $r$ is derived from the burst that is correlated with $b_1$, and that $r'$ is derived from the burst that is correlated with $b_{k-1}$. $Q(b_1|r)$ is obtained using Bayes' theorem~\cite{Joyce2021Bayes} as follows:
\begin{align}
    Q(b_1|r)=\frac{R(r|b_1)Q(b_1)}{R(r)},
\end{align} 
where we derive $R(r|b_1)$ from $Q(b|b_1)$ similarly to $R(r)$ in Eq.~\eqref{eq:R_cont_define}:
\begin{align}
    R(r|b_1) &= \frac{1}{\langle b\rangle}\int_r^\infty {\rm d}b Q(b|b_1)\notag\\
    &=\frac{1}{\langle b\rangle}\int_r^\infty {\rm d}b Q(b) \left[1+\rho f(b)f(b_1) \right]\notag\\
    &=R(r) + \rho f(b_1)R_2(r).
    \label{eq:R(r|b)}
\end{align}
Here we have defined
\begin{align}
    R_2(r)\equiv \frac{1}{\langle b\rangle}\int_{r}^\infty {\rm d}b Q_2(b)
\end{align}
with
\begin{align}
    Q_2(b)\equiv Q(b)f(b).
    \label{eq:Q2_define}
\end{align}
Thus, one gets
\begin{align}
    Q(b_1|r)=Q(b_1)\left[1+\rho \frac{f(b_1) R_2(r)}{R(r)} \right].
    \label{eq:Q_b1|r}
\end{align}
Next, we derive $q(r'|b_{k-1})$ similarly to $q(r')$ in Eq.~\eqref{eq:q_cont_define} as follows:
\begin{align}
    q(r'|b_{k-1})&=\int_{r'}^\infty {\rm d}b Q(b|b_{k-1}) \notag\\
    &= \int_{r'}^\infty {\rm d}b Q(b) \left[1+\rho f(b)f(b_{k-1}) \right] \notag\\
    &=q(r')\left[1+\rho \frac{f(b_{k-1}) q_2(r')}{q(r')} \right],
    \label{eq:q_r|bk1}
\end{align}
where
\begin{align}
    q_2(r')\equiv \int_{r'}^\infty {\rm d}b Q_2(b).
    \label{eq:q2_define}
\end{align}
Similarly, we derive $q(r'|r)$ for the case with $k=1$ in Eq.~\eqref{eq:P[1]_define} as follows:
\begin{align}
    q(r'|r)&=\int_{r'}^\infty {\rm d}b Q(b|r) = \int_{r'}^\infty {\rm d}b\frac{R(r|b)Q(b)}{R(r)}\notag\\
    &=\int_{r'}^\infty {\rm d}b\frac{Q(b)}{R(r)}[R(r) + \rho f(b)R_2(r)]\notag\\
    &=q(r')+\rho\frac{R_2(r)q_2(r')}{R(r)},
    \label{eq:q(r'|r)}
\end{align}
where we have used Eq.~\eqref{eq:R(r|b)}.

For the case with $k=1$, plugging Eq.~\eqref{eq:q(r'|r)} into Eq.~\eqref{eq:P[1]_define}, one obtains
\begin{align}
    P[1]=R(r)q(r')+\rho R_2(r)q_2(r').
    \label{eq:P[1]}
\end{align}
When $k\geq 2$, we plug Eqs.~\eqref{eq:Q_b|b},~\eqref{eq:Q_b1|r},~and~\eqref{eq:q_r|bk1} into Eq.~\eqref{eq:P[k]_define} to obtain
\begin{widetext}
\begin{align}
    P[k] &= R(r) \left[\prod_{i=1}^{k-1}Q(b_i)\right] q(r') 
    \left[1+\rho \frac{f(b_1) R_2(r)}{R(r)} \right]\left\{\prod_{i=1}^{k-2} \left[1+\rho f(b_i)f(b_{i+1})\right]\right\}\left[1+\rho \frac{f(b_{k-1}) q_2(r')}{q(r')} \right] \notag\\
    &\approx R(r) \left[\prod_{i=1}^{k-1}Q(b_i)\right] q(r') \left\{1+\rho\left[\frac{f(b_1) R_2(r)}{R(r)}+\sum_{i=1}^{k-2} f(b_i)f(b_{i+1}) + \frac{f(b_{k-1}) q_2(r')}{q(r')} \right] +\mathcal{O}(\rho^2) \right\}.
    \label{eq:P[k]_approx}
\end{align}    
\end{widetext}
In the second line of Eq.~\eqref{eq:P[k]_approx}, we have expanded the equation up to the first order of $\rho$ by assuming $|\rho|\ll 1$. After plugging Eqs.~\eqref{eq:P[1]} and~\eqref{eq:P[k]_approx} into Eq.~\eqref{eq:p_k_cont} we take the Laplace transformation of $p_k(t_{\rm d})$ to obtain
\begin{align}
    \tilde p_1 =\tilde\psi\tilde R\tilde q +\rho \tilde\psi\tilde R_2\tilde q_2,
    \label{eq:p_1_trans}
\end{align}
for $k=1$ and 
\begin{align}
    \tilde p_k &\approx \tilde\psi^k \tilde Q^{k-1}\tilde R\tilde q +\rho\left[
    \tilde\psi^k \tilde Q^{k-2} \tilde Q_2 (\tilde R_2 \tilde q + \tilde R \tilde q_2)\right. \notag\\
    &\left. + (k-2)\tilde\psi^k \tilde Q^{k-3} \tilde Q_2^2 \tilde R \tilde q\right]+\mathcal{O}(\rho^2),
    \label{eq:p_k_trans}
\end{align}
for $k\geq 2$, respectively. Here we have omitted the Laplace parameter $s$ in all functions for simplicity, and $\tilde\psi$, $\tilde R$, $\tilde q$, $\tilde R_2$, $\tilde q_2$, $\tilde Q$, and $\tilde Q_2$ are respectively Laplace transforms of $\psi(\bar\tau)$, $R(r)$, $q(r')$, $R_2(r)$, $q_2(r')$, $Q(b)$, and $Q_2(b)$. Note that the second term coupled with $\rho$ in Eq.~\eqref{eq:p_k_trans} exists only for $k\geq 3$.

Using Eqs.~\eqref{eq:p_1_trans} and~\eqref{eq:p_k_trans} we now obtain the Laplace transform of $Z(t_{\rm d})$ in Eq.~\eqref{eq:Z_td}:
\begin{align}
    \tilde Z=& \sum_{k=0}^\infty \tilde p_k \approx \tilde Z_0+\rho\tilde Z_1+\mathcal{O}(\rho^2),\label{eq:Zs}
\end{align}
where
\begin{align}
    \tilde Z_0 &\equiv \tilde p_0+\frac{\tilde \psi \tilde R \tilde q}{1-\tilde \psi\tilde Q}=\frac{1}{s} - \frac{(1-\tilde\psi)(1-\tilde Q)}{\langle b\rangle s^2(1-\tilde\psi\tilde Q)},\label{eq:Z0s_alter}\\
    \tilde Z_1 &\equiv \tilde\psi\tilde R_2\tilde q_2+ \frac{\tilde \psi^2 \tilde Q_2(\tilde R_2 \tilde q + \tilde R \tilde q_2)}{1-\tilde \psi\tilde Q} + \frac{\tilde \psi^3 \tilde Q_2^2 \tilde R \tilde q}{(1-\tilde \psi\tilde Q)^2}\notag\\ 
    & = \frac{\tilde\psi (1-\tilde\psi)^2 \tilde Q_2^2}{\langle b\rangle s^2 (1-\tilde\psi\tilde Q)^2}.
    \label{eq:Z1s_alter}
\end{align}
Here the following relations have been used:
\begin{align}
    &\tilde p_0 = \frac{1}{s}-\frac{1-\tilde Q}{\langle b\rangle s^2},\
    \tilde R = \frac{1-\tilde Q}{\langle b\rangle s},\ 
    \tilde q = \frac{1-\tilde Q}{s},\notag\\
    &\tilde R_2 = -\frac{\tilde Q_2}{\langle b\rangle s},\ 
    \tilde q_2 = -\frac{\tilde Q_2}{s}.
\end{align}
We take the inverse Laplace transform of $\tilde Z(s)$ in Eq.~\eqref{eq:Zs} to get $Z(t_{\rm d})$ analytically or numerically. Then, together with $\lambda$ in Eq.~\eqref{eq:lambda}, one can finally derive the analytical solution of the ACF in Eq.~\eqref{eq:acf_simple} up to the first order of $\rho$ for arbitrary reduced IET and burst size distributions. Note that setting $\rho=0$ reduces our solution to the result in Ref.~\cite{Jo2024Temporal}.

\subsection{Exponential case}\label{subsec:exp}

\begin{figure*}[!th]
\centering
\includegraphics[width=0.85\textwidth]{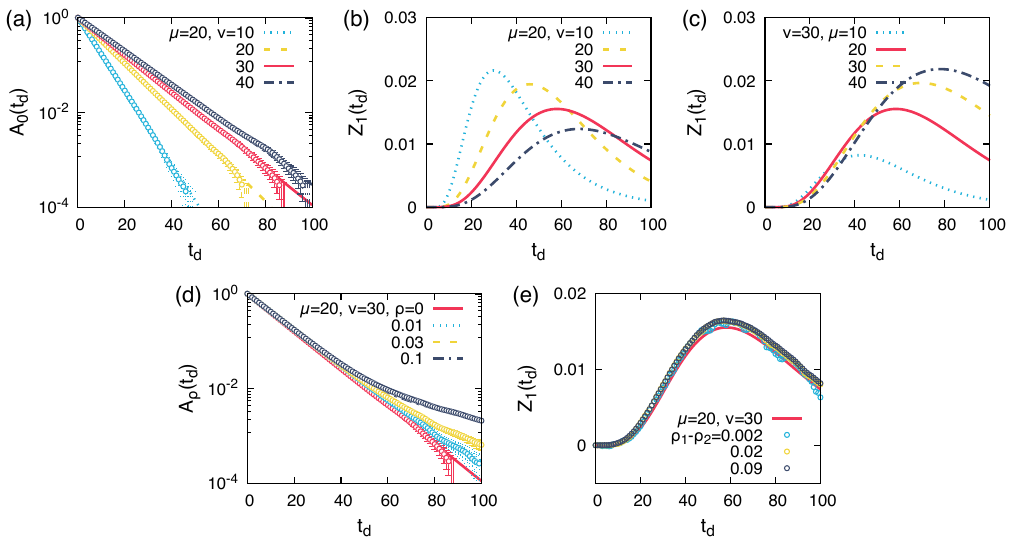}
\caption{(a) Autocorrelation functions in the case of uncorrelated burst sizes, i.e., $A_0(t_{\rm d})$ in Eq.~\eqref{eq:A0td_final} using the numerical inverse Laplace transform of $\tilde Z_0$ in Eq.~\eqref{eq:Z0s_explicit} for $\mu_0=\nu_0=1$, $\mu=20$, and several values of $\nu$ (lines), and their corresponding simulation results, i.e., $\hat A_0(t_{\rm d})$ in Eq.~\eqref{eq:acf_numeric} averaged over $10^3$ different event sequences generated using the same parameter values of $\mu_0$, $\nu_0$, $\mu$, $\nu$, and $T=5\cdot 10^5$ (symbols).
(b,~c) Numerical inverse Laplace transforms of $\tilde Z_1$ in Eq.~\eqref{eq:Z1s_explicit} for $\mu_0=\nu_0=1$, several combinations of $\mu$ and $\nu$.
(d) $A_{\rho}(t_{\rm d})$ in Eq.~\eqref{eq:acf_final} (lines) and $\hat A_\rho(t_{\rm d})$ in Eq.~\eqref{eq:acf_numeric} (symbols) for $\mu_0=\nu_0=1$, $\mu=20$, $\nu=30$, and several values of $\rho$.
(e) Numerical inverse Laplace transform of $\tilde Z_1$ in Eq.~\eqref{eq:Z1s_explicit} for $\mu_0=\nu_0=1$, $\mu=20$, and $\nu=30$ (line) and its corresponding simulation results $\hat Z_1(t_{\rm d})$ in Eq.~\eqref{eq:hatZ1t} for several combinations of $\rho_1$ and $\rho_2$ (symbols). In panels (a,~d), error bars for simulation results stand for the standard errors, while error bars are omitted in the panel (e) for clearer presentation.
}
\label{fig:result}
\end{figure*}

Considering the empirical findings such as heavy-tailed IET and burst size distributions~\cite{Karsai2018Bursty, Jo2020Bursttree}, one could assume the power-law IET and burst size distributions to derive the analytical solution of the ACF as done in Ref.~\cite{Jo2024Temporal}. However, it turns out that the range of $M_b$ in Eq.~\eqref{eq:Mb_arho} becomes extremely narrow when a power-law distribution of $Q(b)$ is used~\cite{Jo2019Analytically}, which is attributed to the fundamental issue of the FGM copula~\cite{Schucany1978Correlation}. In contrast, in the case with the exponential distribution of $Q(b)$, $|M_b|\leq a=1/4$ in Eq.~\eqref{eq:Mb_arho}. Thus, for the demonstration of our analytical result in the previous subsection, we focus on the case with exponential distributions of both reduced IETs and burst sizes with lower bounds of $\mu_0,\nu_0\geq 0$:
\begin{align}
    &\psi(\bar\tau)=\frac{1}{\mu-\mu_0}e^{-\frac{\bar\tau-\mu_0}{\mu-\mu_0}} \theta(\bar\tau-\mu_0),
    \label{eq:psi_mu0}\\
    &Q(b)=\frac{1}{\nu-\nu_0}e^{-\frac{b-\nu_0}{\nu-\nu_0}} \theta(b-\nu_0),
    \label{eq:Q_nu0}
\end{align}
where $\theta(\cdot)$ is a Heaviside step function. It is straightforward to derive that $\langle\bar\tau\rangle=\mu$ and $\langle b\rangle=\nu$. We have introduced lower bounds of $\bar\tau$ and $b$, which is to be consistent with the assumptions that $\bar\tau, b\geq 1$. The case with $\mu_0=\nu_0=0$ has been briefly studied in Ref.~\cite{Jo2024Temporal}. From Eq.~\eqref{eq:Q2_define} one gets
\begin{align}
    Q_2(b)=\frac{1}{\nu-\nu_0}\left(e^{-\frac{b-\nu_0}{\nu-\nu_0}}-2e^{-2\frac{b-\nu_0}{\nu-\nu_0}} \right)\theta(b-\nu_0).
    \label{eq:Q2_nu0}
\end{align}
\begin{widetext}
We obtain the Laplace transforms of Eqs.~\eqref{eq:psi_mu0}--\eqref{eq:Q2_nu0} as follows:
\begin{align}
    \tilde\psi = \frac{e^{-\mu_0 s}}{(\mu-\mu_0)s+1},\ 
    \tilde Q = \frac{e^{-\nu_0 s}}{(\nu-\nu_0)s+1},\ 
    \tilde Q_2 = \frac{-(\nu-\nu_0)s e^{-\nu_0 s}}{\left[(\nu-\nu_0)s+1\right]\left[\nu-\nu_0)s+2\right]},
\end{align}
which are plugged into Eqs.~\eqref{eq:Z0s_alter} and~\eqref{eq:Z1s_alter} to get
\begin{align}
    \label{eq:Z0s_explicit}
    &\tilde Z_0 = \frac{1}{s}-\frac{\left[(\mu-\mu_0)s+1-e^{-\mu_0 s}\right]\left[(\nu-\nu_0)s+1-e^{-\nu_0 s}\right]}{\nu s^2\left[(\mu-\mu_0)(\nu-\nu_0)s^2+(\mu-\mu_0+\nu-\nu_0)s+1-e^{-(\mu_0+\nu_0) s}\right]},\\
    &\tilde Z_1 = \frac{(\nu-\nu_0)^2 \left[(\mu-\mu_0)s+1-e^{-\mu_0 s}\right]^2 e^{-(\mu_0+2\nu_0) s}}{\nu \left[(\mu-\mu_0)s+1\right]\left[(\nu-\nu_0)s+2\right]^2 \left[(\mu-\mu_0)(\nu-\nu_0)s^2+(\mu-\mu_0+\nu-\nu_0)s+1-e^{-(\mu_0+\nu_0) s}\right]^2}.\label{eq:Z1s_explicit}
\end{align}
\end{widetext}
Since it is difficult to take the inverse Laplace transforms of Eqs.~\eqref{eq:Z0s_explicit} and~\eqref{eq:Z1s_explicit} analytically, we instead numerically calculate $Z_0(t_{\rm d})$ and $Z_1(t_{\rm d})$ by using the \texttt{mpmath} library in Python~\cite{Thempmathdevelopmentteam2023Mpmath}. Then we finally get the ACF in Eq.~\eqref{eq:acf_simple}, together with $\lambda=\frac{\nu}{\mu+\nu}$ from Eq.~\eqref{eq:lambda}, as
\begin{align}
    \label{eq:acf_final}
    &A_\rho(t_{\rm d})\approx A_0(t_{\rm d})+\rho\left(\frac{\mu+\nu}{\mu}\right)Z_1(t_{\rm d})+\mathcal{O}(\rho^2),\\
    &A_0(t_{\rm d})\equiv \left(\frac{\mu+\nu}{\mu}\right)\left[Z_0(t_{\rm d})-\frac{\nu}{\mu+\nu}\right],
    \label{eq:A0td_final}
\end{align}
where we have added the subscript $\rho$ to emphasize its role in the result of the ACF.

We first observe the behavior of $A_0(t_{\rm d})$ in Eq.~\eqref{eq:A0td_final}, i.e., in the case with uncorrelated burst sizes, with $\mu_0=\nu_0=1$. Several cases of $A_0(t_{\rm d})$ are depicted as lines in Fig.~\ref{fig:result}(a), showing that $A_0(t_{\rm d})$ overall exponentially decreases with $t_{\rm d}$. The larger average burst size $\nu$ leads to the slower decay of $A_0(t_{\rm d})$. Next, we plot the numerically obtained $Z_1(t_{\rm d})$ for some combinations of $\mu$ and $\nu$ in Fig.~\ref{fig:result}(b,~c); in all cases, we find that as $t_{\rm d}$ increases from $0$, $Z_1(t_{\rm d})$ increases from $0$ and then decreases to eventually approach $0$. Therefore, $Z_1(t_{\rm d})\geq 0$ for the entire range of $t_{\rm d}\geq 0$, implying that the positive correlation between two consecutive burst sizes elevates the ACF curves, as depicted as lines in Fig.~\ref{fig:result}(d). We also observe in Fig.~\ref{fig:result}(b,~c) that for a fixed value of the average reduced IET ($\mu$), the larger value of the average burst size ($\nu$) tends to make the location (height) of the peak of $Z_1(t_{\rm d})$ further from 0 (lower). On the other hand, for a fixed value of $\nu$, the larger value of $\mu$ tends to make the location (height) of the peak of $Z_1(t_{\rm d})$ further from 0 (higher). These results will be compared to the simulation results in the next subsection.

We remark that as the inverse Laplace transforms of $\tilde Z_0$ and $\tilde Z_1$ have been calculated only numerically, one may be interested in the precise functional forms of $Z_0(t_{\rm d})$ and $Z_1(t_{\rm d})$. For this, we assume that $\mu_0=\nu_0=0$, enabling us to explicitly derive the analytical solutions of $Z_0(t_{\rm d})$ and $Z_1(t_{\rm d})$. For the analysis in this case, see Appendix~\ref{append}.

\subsection{Numerical simulations}\label{subsec:numerics}

To demonstrate our analytical results for ACFs in the previous subsection, we generate the event sequence $\{ x(1), \ldots, x(T) \}$ using the following continuous distributions of reduced IETs and burst sizes:
\begin{align}
    &\psi(\bar\tau)=\frac{1}{\mu-1}e^{-(\bar\tau-1)/(\mu-1)} \theta(\bar\tau-1),
    \label{eq:psi_numer}\\
    &Q(b)=\frac{1}{\nu-1}e^{-(b-1)/(\nu-1)} \theta(b-1).
    \label{eq:Q_numer}
\end{align}
Whenever we draw values, i.e., $\bar\tau$ or $b$, from the above distributions, we take their rounded integers, namely, $\bar\tau'\equiv \lfloor \bar\tau+1/2\rfloor$ or $b'\equiv \lfloor b+1/2\rfloor$, where $\lfloor\cdot\rfloor$ is a floor function. 

To be precise, we first draw a burst size $b_1$ from $Q(b)$ in Eq.~\eqref{eq:Q_numer} to set $x(t)=1$ for $t=1,\ldots,b'_1$. Then we randomly draw a reduced IET $\bar\tau_1$ from $\psi(\bar\tau)$ in Eq.~\eqref{eq:psi_numer} to set $x(t)=0$ for $t=b'_1+1,\ldots,b'_1+\bar\tau'_1$. We draw a burst size $b_2$ from $Q(b|b_1)$ in Eq.~\eqref{eq:Q_b|b}, for which we use the numerical method proposed in Ref.~\cite{Jo2019Copulabased}, to set $x(t)=1$ for $t=b'_1+\bar\tau'_1+1,\ldots,b'_1+\bar\tau'_1+b'_2$. Another reduced IET $\bar\tau_2$ is drawn from $\psi(\bar\tau)$ to set $x(t)=0$ for $t=b'_1+ \bar\tau'_1 +b'_2+1, \ldots, b'_1 +\bar\tau'_1 +b'_2+ \bar\tau'_2$. This procedure is repeated until $t$ becomes $T$. 

Once the time series $\{ x(1), \ldots, x(T) \}$ is generated, we numerically calculate the ACF by
\begin{align}
    \hat A_\rho(t_{\rm d})\equiv \frac{\frac{1}{T-t_{\rm d}} \sum_{t=1}^{T-t_{\rm d}} x(t)x(t+t_{\rm d})-\lambda_0\lambda_1}{\sigma_0\sigma_1},
    \label{eq:acf_numeric}
\end{align}
where $\lambda_0$ and $\sigma_0$ are the average and standard deviation of $\{x(1), \ldots, x(T-t_{\rm d})\}$, and $\lambda_1$ and $\sigma_1$ are the average and standard deviation of $\{x(t_{\rm d}+1),\ldots, x(T)\}$. For the simulation, we generate $10^3$ different event sequences with $T=5\cdot 10^5$ and $\mu_0=\nu_0=1$ for several combinations of $\mu$, $\nu$, and $\rho$. Then we calculate $\hat A_\rho(t_{\rm d})$ from each event sequence and take the average of them. The simulation results of $\hat A_\rho(t_{\rm d})$ are plotted in Fig.~\ref{fig:result}(a,~d), which appear to be in good agreement with the analytical solutions in Eq.~\eqref{eq:acf_final}. 

In order to isolate the effect of the correlation parameter $\rho$ on the simulation results of the ACF, we define the following difference between two ACFs using the same parameter values but with different $\rho$ values:
\begin{align}
    \hat Z_1(t_{\rm d})\equiv \frac{\mu}{\mu+\nu}\frac{\left[\hat A_{\rho_1}(t_{\rm d})-\hat A_{\rho_2}(t_{\rm d})\right]}{(\rho_1-\rho_2)}.
    \label{eq:hatZ1t}
\end{align}
From Eq.~\eqref{eq:acf_final}, it is obvious why the above quantity in Eq.~\eqref{eq:hatZ1t} is denoted by the same symbol as $Z_1(t_{\rm d})$. As expected, the results of $\hat Z_1(t_{\rm d})$ for several combinations of $\rho_1$ and $\rho_2$ are in good agreement with the analytical solutions, as depicted in Fig.~\ref{fig:result}(e). Some systematic discrepancy between $\hat Z_1(t_{\rm d})$ and $Z_1(t_{\rm d})$ might be due to the higher-order terms of $\rho$ that have been ignored in the derivation of the analytical solution.

\section{Conclusion}

To study the effect of the correlation between two consecutive burst sizes on the autocorrelation function (ACF) of the time series, we derive the analytical solution of the ACF for arbitrary interevent time (IET) and burst size distributions. The correlations between two consecutive burst sizes are implemented using the Farlie-Gumbel-Morgenstern (FGM) copula, enabling us to control such correlations using a single correlation parameter $\rho$. Assuming exponential distributions of IETs and burst sizes with lower bounds, we analytically and numerically find that the positive correlation between two consecutive burst sizes elevates the ACF curve for the intermediate range of the time lag. 

For the analytical tractability, we have assumed that the IETs and burst sizes are continuous variables, although the proposed model is originally defined in discrete times. Therefore, the analysis of the ACF using the discrete version of FGM copula~\cite{Piperigou2009Discrete} might be more accurate than our current analysis, which is left for a future work. In addition, heavy-tailed IET and burst size distributions might be employed for more realistic modeling. Regarding this, our results can serve as a reference to study more realistic models for the time series showing higher-order temporal correlations beyond the IET distribution.

\begin{acknowledgments}
H.-H.J. acknowledges financial support by the National Research Foundation of Korea (NRF) grant funded by the Korea government (MSIT) (No. 2022R1A2C1007358).
\end{acknowledgments}

\appendix 

\section{Analytical solutions for $\mu_0=\nu_0=0$}\label{append}

Here we analytically study the case with $\mu_0=\nu_0=0$. By setting $\mu_0=\nu_0=0$ in Eqs.~\eqref{eq:Z0s_explicit} and~\eqref{eq:Z1s_explicit}, one gets
\begin{align}
    \label{eq:Z0s_exp}
    \tilde Z_0 &\equiv \frac{\nu}{(\mu+\nu)s} + \frac{\mu}{(\mu+\nu)\left(s+\frac{\mu+\nu}{\mu\nu}\right)},\\
    \tilde Z_1 &\equiv 
    \frac{1}{\mu\nu^3\left(s+\tfrac{1}{\mu}\right)\left(s+\tfrac{2}{\nu}\right)^2\left(s+\tfrac{\mu+\nu}{\mu\nu}\right)^2}.
    \label{eq:Z1s_exp}
\end{align}
Taking the inverse Laplace transform of $\tilde Z_0$ in Eq.~\eqref{eq:Z0s_exp}, we obtain
\begin{align}
    Z_0(t_{\rm d}) = \frac{\nu}{\mu+\nu} + \frac{\mu}{\mu+\nu} e^{-t_{\rm d}/t_0},\ t_0 \equiv \frac{\mu\nu}{\mu+\nu}.
    \label{eq:Z0td}
\end{align}
When $\mu\neq\nu$ and $2\mu\neq \nu$, we take the inverse Laplace transform of $\tilde Z_1$ in Eq.~\eqref{eq:Z1s_exp} to obtain
\begin{align}
    Z_1(t_{\rm d})&=(c_0+c'_0 t_{\rm d})e^{-t_{\rm d}/t_0} + c_\mu e^{-t_{\rm d}/\mu} 
    \notag\\
    & +(c_{\nu}+c'_{\nu}t_{\rm d})e^{-2t_{\rm d}/\nu},
    \label{eq:Z1td}
\end{align}
where
\begin{align}
    &c_0 \equiv \frac{\mu\nu(\mu+\nu)}{(\mu-\nu)^3},\
    c'_0 \equiv \frac{-\mu}{(\mu-\nu)^2},\
    c_\mu\equiv \frac{\mu\nu}{(2\mu-\nu)^2}, \notag\\
    &c_{\nu}\equiv \frac{\mu^3\nu(-5\mu+3\nu)}{(\mu-\nu)^3(2\mu-\nu)^2},\ 
    c'_{\nu}\equiv \frac{-\mu^2}{(\mu-\nu)^2(2\mu-\nu)}.\notag
\end{align}
Note that $c_0+c_\mu+c_\nu=0$, implying that $Z_1(0)=0$. Combining Eqs.~\eqref{eq:Z0td}~and~\eqref{eq:Z1td}, together with $\lambda=\frac{\nu}{\mu+\nu}$ from Eq.~\eqref{eq:lambda}, we finally derive the analytical solution of the ACF in Eq.~\eqref{eq:acf_simple} up to the first order of $\rho$:
\begin{align}
    A(t_{\rm d}) &\approx e^{-t_{\rm d}/t_0} + \rho\left(\frac{\mu+\nu}{\mu}\right)\left[(c_0+c'_0t_{\rm d})e^{-t_{\rm d}/t_0}  \right. \notag\\
    &\left. + c_\mu e^{-t_{\rm d}/\mu}
    + (c_{\nu}+c'_{\nu}t_{\rm d})e^{-2t_{\rm d}/\nu}\right]+\mathcal{O}(\rho^2).
    \label{eq:Atd_final}
\end{align}
 
We also calculate $Z_1(t_{\rm d})$ and $A(t_{\rm d})$ for the special case with $\mu=\nu$ as
\begin{align}
    Z_1(t_{\rm d}) &\approx
    e^{-t_{\rm d}/\nu} - \left(1 + \frac{t_{\rm d}}{\nu}+ \frac{t_{\rm d}^2}{2\nu^2} + \frac{t_{\rm d}^3}{6\nu^3} \right)e^{-2t_{\rm d}/\nu}, 
    \label{eq:Z1td_final_munu} \\ 
    A(t_{\rm d}) &\approx e^{-2t_{\rm d}/\nu} + 2\rho\left[e^{-t_{\rm d}/\nu}
    \right.\notag\\
    &\left.- \left(1 + \frac{t_{\rm d}}{\nu}+ \frac{t_{\rm d}^2}{2\nu^2} + \frac{t_{\rm d}^3}{6\nu^3} \right)e^{-2t_{\rm d}/\nu}
    \right]+\mathcal{O}(\rho^2),
    \label{eq:Atd_final_munu}
\end{align}
where we have used the fact that $\lambda=1/2$. In the case with $2\mu=\nu$, we get using $\lambda=2/3$
\begin{align}
    Z_1(t_{\rm d}) &\approx
    \left(6 - \frac{4t_{\rm d}}{\nu}+\frac{t_{\rm d}^2}{\nu^2}\right)e^{-2t_{\rm d}/\nu} -\left(6 + \frac{2t_{\rm d}}{\nu}\right)e^{-3t_{\rm d}/\nu},
    \label{eq:Z1td_final_munu2} \\ 
    A(t_{\rm d}) &\approx e^{-3t_{\rm d}/\nu} + 3\rho\left[\left(6 - \frac{4t_{\rm d}}{\nu}+\frac{t_{\rm d}^2}{\nu^2}\right)e^{-2t_{\rm d}/\nu}
    \right.\notag\\
    &\left.
    -\left(6 + \frac{2t_{\rm d}}{\nu}\right)e^{-3t_{\rm d}/\nu}
    \right]+\mathcal{O}(\rho^2).
    \label{eq:Atd_final_munu2}
\end{align}
In both cases of $\mu=\nu$ and $2\mu=\nu$, one obtains $Z_1(0)=0$. All these analytical solutions are qualitatively similar to the results in the case of $\mu_0,\nu_0>0$ in the main text.

%


\end{document}